\documentclass[pra,twocolumn,showpacs,superscriptaddress]{revtex4-1}

\usepackage{placeins}
\usepackage{amssymb}
\usepackage{amsmath}
\usepackage{pstricks}
\usepackage{epsfig}
\newcommand{\ket}[1]{\left| #1 \right\rangle}

\newcommand{\fr}{{\rm f}}
\newcommand{\brnew}{{\rm b}}
\newcommand{\rmi}{{\rm i}}
\newcommand{\rmd}{\mathrm{d}}

\newcommand{\ie}{{\it{i.e.}}}
\newcommand{\refeq}[1]{equation (\ref{#1})}
\newcommand{\refeqs}[1]{equations (\ref{#1})}

\begin{document}
\title{Optical quantum memory for polarization qubits with $V$-type three-level atoms}

\author{D.~Viscor}
\address{Departament de F\'{\i}sica, 
Universitat Aut\`{o}noma de Barcelona, E-08193 Bellaterra, Spain} 

\author{A.~Ferraro}
\address{Departament de F\'{\i}sica, 
Universitat Aut\`{o}noma de Barcelona, E-08193 Bellaterra, Spain} 

\author{Yu.~Loiko}
\address{Departament de F\'{\i}sica, 
Universitat Aut\`{o}noma de Barcelona, E-08193 Bellaterra, Spain} 
\address{Institute of Physics,
National Academy of Sciences of Belarus, Nezalezhnasty Ave. 68, 220072 Minsk, Belarus}

\author{R.~Corbal\'an}
\address{Departament de F\'{\i}sica, 
Universitat Aut\`{o}noma de Barcelona, E-08193 Bellaterra, Spain} 

\author{J.~Mompart}
\address{Departament de F\'{\i}sica, 
Universitat Aut\`{o}noma de Barcelona, E-08193 Bellaterra, Spain} 

\author{V.~Ahufinger}
\address{Departament de F\'{\i}sica, 
Universitat Aut\`{o}noma de Barcelona, E-08193 Bellaterra, Spain} 
\address{ICREA - Instituci\'{o} Catalana de Recerca i Estudis Avan\c{c}ats, Barcelona, Spain }

\begin{abstract}

We investigate an optical quantum memory scheme with $V$-type three-level atoms based on the controlled reversible inhomogeneous broadening (CRIB) technique.
We theoretically show the possibility to store and retrieve a weak light pulse interacting with the two optical transitions of the system.
This scheme implements a quantum memory for a polarization qubit
--- a single photon in an arbitrary polarization state ---
without the need of two spatially separated two-level media,
thus offering the advantage of experimental compactness overcoming the limitations due to mismatching and unequal efficiencies that can arise in spatially separated memories.
The effects of a relative phase change between the atomic levels,
as well as of phase noise due to, for example, the presence of spurious electric and magnetic fields are analyzed.

\end{abstract}

\date{\today}



\maketitle

\section{\label{sec:intro}Introduction}

Optical quantum memories aim at the storage and retrieval of quantum states of light by means of physical systems with long-living coherences. The realization of high efficient quantum memories is an ambitious goal towards which there is a considerable worldwide activity \cite{nature_review,Zhao'09,Riedmatten'10,Reim'10,Saglamyurek'11,Clausen'11,LongdellNews'11}. Optical quantum memories are essential for quantum information processing: they play a central role in quantum computation with linear optics \cite{klm} and, more generally, in quantum networks \cite{kimble}. Specifically, optical quantum communication suffers from degradation of information along transmission at large distances. To cope with this, advanced strategies have been proposed requiring the use of quantum repeaters of which quantum memories are the core ingredient \cite{qrep}.

In quantum communication with photons, logical qubits can be encoded in several ways, for example via polarization, time-bin, path, phase, or photon-number encodings. Concerning polarization, the logical $\ket{0}$ and $\ket{1}$ quantum states correspond to two orthogonal polarization degrees of freedom, for example, left- and right- circular polarizations. It is on the storage and retrieval of polarization qubits that we will focus on this work. Up to now, polarization-qubit memories have been experimentally realized mainly in cold atomic ensembles and atomic vapors \cite{polmem_ensembles}. In these experiments, polarization encoding is first transformed into path encoding and then stored into two spatially distinguishable ensembles (actually, two different memories). Retrieval is eventually performed by mixing again the output pulses, thus transforming back to the original polarization encoding.

The above mentioned realizations are attractive, however they suffer from several practical drawbacks due to the extra step of splitting spatially the input state into two beams. As a matter of fact, in order to preserve the original input quantum state, the two retrieved beams must be indistinguishable when mixed back again (apart from what concerns their polarization). In particular, they must occupy the same spatial and temporal mode, implying perfect matching at the mixing stage. In addition, the efficiency of the two memories must be the same, otherwise the two polarization components of the original beam would be retrieved in an unbalanced fashion. Such necessary requirements are experimentally demanding. 
Clearly, similar problems would affect also solid-state based memories, if two different solid-state memories were used to store each polarization component along the lines of~\cite{polmem_solid}.
Recently, the storage of polarization qubits without spatially splitting the input photon has been addressed~\cite{Specht'11,Carreño'10}. In~\cite{Specht'11} the qubit is stored in a single atom inside a cavity by means of stimulated Raman adiabatic passage whereas in~\cite{Carreño'10} a dense ensemble of four-level atoms in a tripod configuration is investigated for storage and retrieval of polarization qubits.

We show here that a quantum memory for polarization qubits can be implemented also in a simple three-level system without the extra step of splitting spatially the input state. In particular we base our analysis in the photon echo technique \cite{peqm_review}. The latter has been developed especially for the storage of time-bin qubits, being particularly suited for broad-band pulses, in contrast with the proposal in~\cite{Specht'11}. However, we will show in this work that photon echo quantum memories could also offer a compact resource to store and retrieve polarization qubits.

Based on photon echo processes, two main approaches have been developed for devising quantum memories in solids: controlled reversible inhomogeneous broadening (CRIB) \cite{MK,crib,Kraus'06,SangouardPRA07,Longdell'08,GEM,Sparkes'10,HetetOptLett'08,Hosseini'09} and atomic frequency comb (AFC) quantum memories \cite{Saglamyurek'11,Clausen'11,afc}. Here, we will mainly focus on the former, even if adaptation to AFC can be envisaged with a similar approach. Quantum memories based on CRIB were first proposed in a seminal article by Moiseev and Kr\"{o}ll \cite{MK}, where an optical pulse was stored and retrieved in a Doppler broadened two-level atomic vapor. 
In this technique, since the medium is inhomogeneously broadened, the pulse is absorbed collectively by all the atoms in a coherent way, so each one evolves with a different phase depending on its detuning. In this scheme, an auxiliary $\pi$ pulse is used, after the absorption, to increase the storage time by transferring the atomic coherences to a long-living third level. Afterwards, another counter-propagating $\pi$ pulse brings the population back to the original two-level transition and 
imprints into the medium the phase matching conditions 
needed for the retrieval of the light pulse in the backward direction.
As an automatic consequence of the Doppler effect,
the inhomogeneous broadening of the original two-level transition is now reversed for the backward propagating retrieved pulse.
Thus, a backward propagating time-reversed replica of the original pulse is retrieved.
Further, the CRIB technique has been extended to solid-state materials \cite{crib}.
The key point is that in appropriate materials the relevant two-level transition can be artificially broadened, and the broadening reversed in a controlled way after the absorption and storage of the incident pulse.
In later contributions \cite{SangouardPRA07,Longdell'08} it was shown that, if the phase-matching operation is not performed, the retrieved pulse propagates in the forward direction. In this case, the re-absorption by the medium itself limits the output efficiency to $\sim54\%$. 
An alternative to this method is the so-called longitudinal CRIB or gradient-echo memory (GEM) \cite{GEM}, in which the artificial inhomogeneous broadening is not created at each spatial point, but it is created in such a way that the detuning changes linearly along the propagation direction of the light. Thus, each frequency component of the incident pulse is absorbed at a different position inside the medium, and once the detuning is reversed, this allow for a perfect retrieval efficiency even in the forward direction for large optical depths. In this work we focus in the original proposal, called the transverse CRIB in contraposition to the longitudinal case, although the results shown here could be easily extended to the GEM scheme.

In most of the previous works concerning CRIB, usually only two-level systems have been considered ---apart from the mentioned auxiliary third level used only to perform the phase-matching operation and increase the storage time. Here we extend the CRIB approach to a medium composed of three-level atoms, paying special attention to the influence of the phase between the different atomic levels in the storage and retrieval efficiencies. We consider a collection of inhomogeneously broadened $V$-type three-level atoms (see figure \ref{f:figs1ab}), initially prepared in their ground state. 
The medium interacts with a weak pulse with two orthogonal polarization components, that is to say, left- and right-circular polarizations, each one coupled with one of the two optical transitions. With this scheme we will show that a $V$-type three-level medium can be used as a quantum memory for single photons in an arbitrary superposition of polarization states, without the need of spatially splitting the input pulse.

The paper is organized as follows. In Section~\ref{sec:Model} we will present the physical system under investigation and derive the optical-Bloch equations that govern its dynamical evolution. We will also introduce the figures of merit that will allow us to assess the quality of the quantum memory protocol, namely the efficiency and the fidelity of the retrieval process. In Section~\ref{sec:ideal case} we will report the analytical solutions for the evolution equations of the system at the stages of pulse storage and retrieval for both backward and forward modes. We will show that under the weak excitation regime, the system is equivalent to two independent two-level quantum memories. 
Next, in Section~\ref{sec:PhaseNoise} we will analyze the effects ---on both the efficiency and the fidelity of the quantum memory--- of a relative phase change between the different atomic levels as well as of phase noise during the storage time. In Section~\ref{sec:Numerics} we will verify the analytical predictions by the numerical integration of the full set of optical-Bloch equations. Finally, in Section~\ref{sec:Conclusions}, we will summarize the results and conclude.

\section{\label{sec:Model}Basic Equations of CRIB for $V$-type three-Level atoms}

\begin{figure}[t]
\begin{center}
\includegraphics[width=0.8\columnwidth]{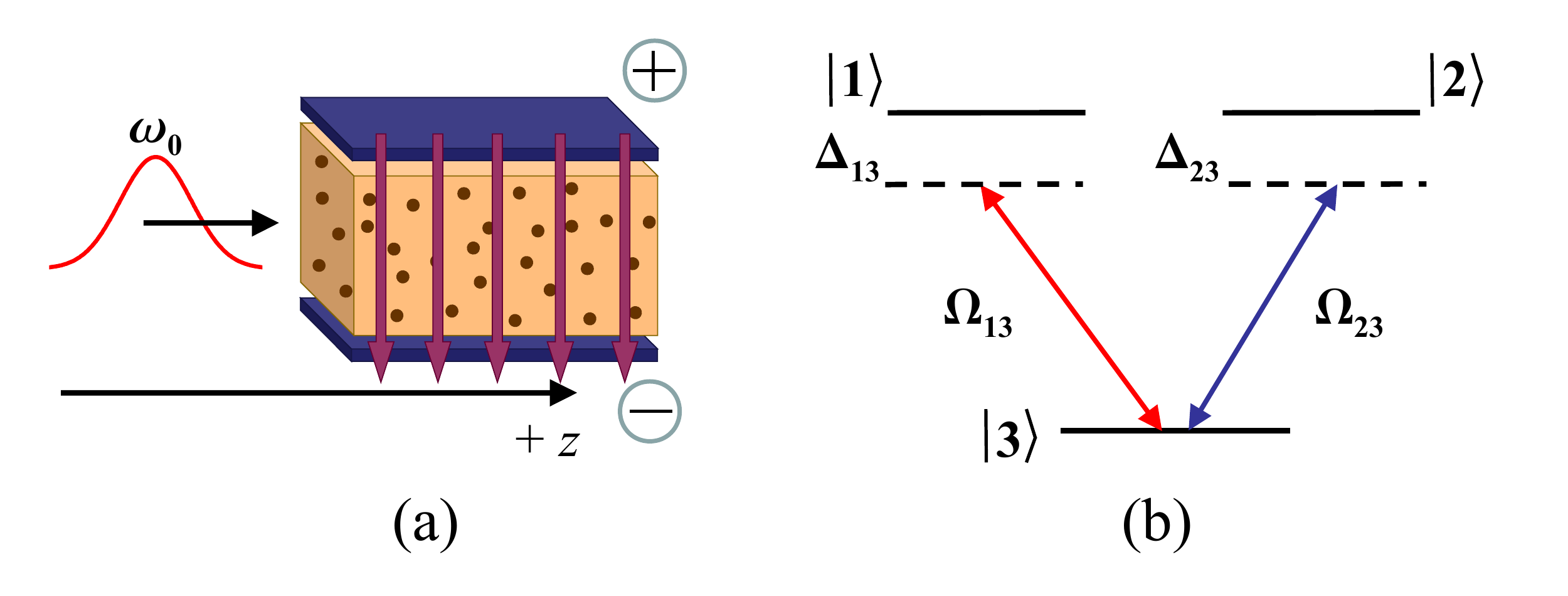}
\end{center}
\caption{(a) Illustration of the physical system under investigation: a single pulse with central frequency $\omega_{0}$ and two circular polarization components enters a medium with transverse inhomogeneous broadening, e.g., by means of a transverse electric field gradient. (b) Scheme of the $V$-type three-level atoms that compose the medium. The left- (right-) circularly polarized component of the incident pulse couples to transition $\left|3\right\rangle \leftrightarrow \left|2\right\rangle$ ($\left|3\right\rangle \leftrightarrow \left|1\right\rangle$), $\Omega_{13}$ and $\Omega_{23}$ denote the corresponding Rabi frequencies and $\Delta_{13}$ and $\Delta_{23}$ the detunings.
}
\label{f:figs1ab}
\end{figure}

\FloatBarrier
The scenario we address here is sketched in figure \ref{f:figs1ab}. 
A single arbitrarily polarized weak pulse
propagates in the forward direction ($+z$) and interacts with a medium composed of $V$-type three-level atoms,
whose transitions have been subjected to artificial 
transverse inhomogeneous broadening (figure \ref{f:figs1ab}(a)), induced for example with a transverse electric field gradient. Note that in order to have a complete absorption of the field the inhomogeneous broadening width must be larger than the spectral width of the input pulse. 
 We assume the same nominal frequencies for the two optical transitions of the $V$-scheme and that the left- and right- circular polarization components of the weak pulse interact with a particular optical transition (figure \ref{f:figs1ab}(b)).

In this situation, the temporal evolution of a single atom 
can be described by the following density matrix equations:
\begin{subequations} \label{sigmaijeqs}
\begin{eqnarray}
\frac{\partial}{\partial t}\sigma_{11}(z,t)&=& \rmi\sigma_{13}(z,t)\Omega_{13}^{\ast}(z,t)+ {\rm c.c.} \\
\frac{\partial}{\partial t}\sigma_{22}(z,t)&=& \rmi\sigma_{23}(z,t)\Omega_{23}^{\ast}(z,t)+ {\rm c.c.} \\
\frac{\partial}{\partial t}\sigma_{12}(z,t)&=& \rmi\omega_{12}\sigma_{12}(z,t)+\rmi\sigma_{13}(z,t)\Omega_{23}^{\ast}(z,t) \notag \\
&&-\rmi\sigma_{32}(z,t)\Omega_{13}(z,t) \label{sigma12eqs} \\
\frac{\partial}{\partial t}\sigma_{13}(z,t)&=& \rmi\omega_{13}\sigma_{13}(z,t) +\rmi\sigma_{12}(z,t)\Omega_{23}(z,t) \notag \\
&&-\rmi\left[\sigma_{33}(z,t)-\sigma_{11}(z,t)\right]\Omega_{13}(z,t) \label{sigma13eqs} \\
\frac{\partial}{\partial t}\sigma_{23}(z,t)&=& \rmi\omega_{23}\sigma_{23}(z,t) +\rmi\sigma_{21}(z,t)\Omega_{13}(z,t) \notag \\
&&-\rmi\left[\sigma_{33}(z,t)-\sigma_{22}(z,t)\right]\Omega_{23}(z,t) \label{sigma23eqs}
\end{eqnarray}
\end{subequations}
where $\sigma_{ii}$ denotes the population of level $\left|i \right\rangle$, $\sigma_{ij}$ is the atomic coherence between levels $\left|i \right\rangle$ and $\left|j \right\rangle$, $\Omega_{ij}=\left( \vec{d}_{ij}\vec{E}_{ij} \right) /\hbar$ 
is the Rabi frequency with $\vec{E}_{ij}$ 
being the slowly varying electric field amplitude of the light component coupled with transition $\left|i\right\rangle \leftrightarrow \left|j\right\rangle$, $\vec{d}_{ij}$ is the electric dipole moment of the corresponding transition,
$\hbar$ is the reduced Planck constant,
and $\omega_{ij}=\omega_{i}-\omega_{j}$ is the transition frequency between levels $\left|i \right\rangle$ and $\left|j \right\rangle$.
Note that, since we consider a closed system, the atomic populations must satisfy $\sigma_{11}+\sigma_{22}+\sigma_{33}=1$.
In addition, for simplicity, we have not included any incoherent decay term in equations (\ref{sigmaijeqs}).
This is well justified since we restrict ourselves to the case for which the lifetimes of the excited levels
and of the optical coherences are much longer than the total time of the storing and retrieving processes. 
Note that this is a common assumption done in the literature. However, spontaneous emission is not the unique decoherence mechanism that could affect the efficiency or the fidelity of the quantum memory implementation. In particular, in Section~\ref{sec:PhaseNoise} we consider different phase noise mechanisms affecting the atomic transitions due to, for instance, the transfer of the stored excitation back and forth from the excited state to the metastable one, or an imperfect isolation of the atomic ensemble from external electric or magnetic fields.

In order to solve analytically equations (\ref{sigmaijeqs}), we generalize the treatment detailed in~\cite{SangouardPRA07} to the case of a three-level medium.
With this aim, we split the operators associated with the two dipole transitions as well as the Rabi frequencies of the two field components in forward and backward modes, denoted by
the superscripts ``f'' and ``b'', respectively:
\begin{subequations}
\begin{eqnarray} 
\sigma_{\mu\rho}(z,t)&=&\sigma_{\mu\rho}^{\fr}(z,t)\mathrm{e}^{\rmi(\omega_{0}t-k_{0}z)}+\sigma_{\mu\rho}^{\brnew}(z,t)\mathrm{e}^{\rmi(\omega_{0}t+k_{0}z)} \label{sigmaf-b} \notag \\ \\
\Omega_{\mu\rho}(z,t)&=&\Omega_{\mu\rho}^{\fr}(z,t)\mathrm{e}^{\rmi(\omega_{0}t-k_{0}z)}+\Omega_{\mu\rho}^{\brnew}(z,t)\mathrm{e}^{\rmi(\omega_{0}t+k_{0}z)}  \label{fieldf-b} \notag \\
\end{eqnarray}
\end{subequations}
where $\omega_0$ is the central frequency of the pulse and $k_0$ the corresponding wavenumber. From now on, we set $\rho=3$ and $\mu=1,2$. For a weak input pulse the dynamical evolution of the level populations can be neglected, and therefore, since the initial population is assumed to be in the ground level $\left\vert 3\right\rangle$, one can assume that $\sigma_{33}\simeq1$ and $\sigma_{11}=\sigma_{22}=\sigma_{12}\simeq0$ during the whole process. With these assumptions, and decomposing the field into forward and backward modes, equations (\ref{sigmaijeqs}) can be written as follows:
\begin{eqnarray} \label{sigma13eq}
\frac{\partial}{\partial t}\sigma_{\mu\rho}^{\brnew,\fr}(z,t,\Delta_{\mu\rho})&=& \rmi\Delta_{\mu\rho}\sigma_{\mu\rho}^{\brnew,\fr}(z,t,\Delta_{\mu\rho})-\rmi\Omega^{\brnew,\fr}_{\mu\rho}(z,t) \notag\\ &&+\rmi\sigma_{\mu\nu}\Omega^{\brnew,\fr}_{\nu\rho}(z,t),
\end{eqnarray}
where we have made explicit the dependence of the optical coherence on the detuning $\Delta\equiv\omega_{13}-\omega_{0}=\omega_{23}-\omega_{0}$, and $\rho=3$, $\mu,\nu=1,2$ with $\mu\neq\nu$. 

The propagation of the forward and backward modes of light,
in a reference frame moving with the pulses
($z \rightarrow z$ and $t \rightarrow t \mp z/c$), is given by the following equations:
\begin{eqnarray} \label{fieldeqs}
\frac{\partial}{\partial z}\Omega^{\brnew,\fr}_{\mu\rho}(z,t) &=&\mp \rmi\eta_{\mu\rho}\int^{\infty}_{-\infty}{ G_{\mu\rho}\left(\Delta\right)\sigma_{\mu\rho}^{\brnew,\fr}(z,t,\Delta)\rmd\Delta}\;, \label{fieldeqf} \notag \\
\end{eqnarray}
where $-$ ($+$) refers to the backward (forward) mode, $G_{\mu\rho}\left(\Delta\right)$ are the inhomogeneous atomic frequency distributions and $\eta_{\mu\rho}=g^{2}Nd^{2}_{\mu\rho}/\hbar c$ with $g^{2}=\omega_{0}/2\varepsilon_{0}V$ being the coupling constant, $\varepsilon_{0}$ the vacuum permittivity, $N$ the number of atoms in the quantization volume $V$,
and $c$ is the speed of light in vacuum.
It is worth noting that, although the treatment is performed in the semi-classical formalism, the linearity of the equations ensures the validity of this model also at the single photon level. Thus, in what follows, both the atomic coherences and the fields could be interpreted as classical amplitudes as well as quantum operators.

From (\ref{sigma13eq}) and (\ref{fieldeqs}), it is direct to see that the time reversed ($t\rightarrow-t$) equations for the forward propagating modes are identical to the backward ones under a sign change in the detunings and in the field amplitudes.
This symmetry in the optical-Bloch equations is indeed the basis for the CRIB protocol.
Thus, once the forward propagating input light pulse has been completely absorbed,
the reversing of the detunings ($\Delta\rightarrow-\Delta$)
should allow to retrieve a time reversed copy of the input pulse.
This reversing operation can be achieved, for instance, by changing the polarity of the device that creates the inhomogeneous broadening. 
At the same time, one has to apply a position dependent phase matching operation to transform the forward components of the atomic excitations into backward components, such that the retrieved field propagates in the backward direction. This phase matching operation can be performed, for instance, by applying two counter-propagating $\pi$ pulses to transfer the atomic coherences back and forth to an auxiliary metastable state \cite{crib,Kraus'06}, which also
allows for longer storage times \cite{spinCoh}. If the phase matching is not performed,
the atoms will reemit in the forward direction, process that is described by the forward modes of equations (\ref{sigma13eq}) and (\ref{fieldeqs}).

In this work we study both the backward and forward retrieval protocols, whose quality is assessed by defining the retrieval efficiency and fidelity as follows.
For each of the two field components involved in the process, the efficiency is defined as:
\begin{eqnarray} \label{def_eff}
	{\rm Eff}^{\brnew,\fr}_{\mu\rho}&=&\frac{\int \rmd\omega{\left| \widetilde{\Omega}_{\mu\rho}^{\brnew,\fr}(z=z_{\rm out},\omega)\right|^{2}}}{\int \rmd\omega{\left| \widetilde{\Omega}^{\rm in}_{\mu\rho}(z=0,\omega)\right|^{2}}},
\end{eqnarray}
where the superscripts $\fr$ and $\brnew$ refer to the forward and backward protocols, and $\widetilde{\Omega}^{\rm in}_{\mu\rho}(0,\omega)$ and $\widetilde{\Omega}_{\mu\rho}^{\brnew,\fr}(z_{\rm out},\omega)$
denote the temporal 
Fourier transforms of the input and output fields, respectively.
For the backward and forward retrieval schemes
in a medium with length $L$,
the output fields are evaluated at the medium surfaces
$z_{\rm out}=0$ and $z_{\rm out}=L$, respectively.
Equation (\ref{def_eff}) addresses
the energetic balance of the storage and retrieval processes, that is to say, the storage and retrieval of each component of the field independently from the other one. However, when considering the storage of quantum fields, it is of relevance to focus on the phase between the two components (polarizations) of the retrieved qubit by comparing it with the phase of the input qubit. 
Since we are interested in the case of initial pure states, the fidelity is defined as the overlap between the initial state $| \psi_{\rm in} \rangle$ and the retrieved state, which in general, could be pure or mixed:
\begin{eqnarray} \label{def_fid}
	{\rm F}^{\brnew,\fr}&=&|\langle \psi_{\rm in} | \varrho^{\brnew,\fr} |\psi_{\rm in}\rangle|^2,
\end{eqnarray}
where $\varrho^{\brnew,\fr}$ is the density matrix of the backward and forward retrieved qubits. Note that for those cases in which the quantum memory is perfectly efficient the retrieved state is pure. This is due to the unitarity of the dynamics and the fact that the medium and the fields end up in a factorized state. On the contrary, for a non-unit efficiency, for instance for finite optical depths, where part of the incident light is lost, the output state will be mixed.

\section{\label{sec:ideal case}Quantum storage in $V$-type three-level atoms}

In this Section we solve analytically equations (\ref{sigma13eq}) and (\ref{fieldeqs}), obtained under the weak input pulse approximation and considering that the atomic population is initially in the ground state.
Under the weak field approximation, the population of the excited levels as well as the coherence between them remains negligible during the whole storage and retrieval process. In such case the evolution of the two components of the pulse are uncoupled from each other
and equations (\ref{sigma13eq}) can be solved for each field component separately.
In other words, the system turns out to be equivalent to two independent two-level media.
Thus, the standard treatment of CRIB in two-level media can be applied:
each pulse component is absorbed by the transition it is coupled with,
and the information is stored in the respective optical atomic coherence.
Afterwards, each pulse can be retrieved as in standard CRIB, 
either in the backward or in the forward direction. Thus, this protocol avoids the need of any active spatial 
separation of the pulses to store them. Therefore, the use of CRIB in $V$-type three level atoms does not suffer from detrimental effects due to the mismatch in spatially separated memories. For later purposes, let us now 
present the solution for the field in all the stages of the CRIB protocol following the lines of~\cite{SangouardPRA07}.

\subsection{\label{ssec: absorption} Absorption}

Within the weak input pulse approximation and, therefore, neglecting the dynamics of the two-photon coherence $\sigma_{12}$, the equations for the two field components decouple and equations (\ref{sigma13eq})-(\ref{fieldeqs}) can be analytically solved inserting the solution of equations (\ref{sigma13eq}) into equations (\ref{fieldeqs}). Using the notation introduced in equation (\ref{def_eff}) and Fourier transforming the result, one obtains: 
\begin{eqnarray}  \label{fieldeqijfin}
\frac{\partial}{\partial z}\widetilde{\Omega}^{\rm in}_{\mu\rho}(z,\omega) &=&-\eta_{\mu\rho}H_{\mu\rho}(\omega)\widetilde{\Omega}^{\rm in}_{\mu\rho}(z,\omega),
\end{eqnarray}
where we have defined
\begin{eqnarray}
H_{\mu\rho}(\omega)=\int^{\infty}_{-\infty} G_{\mu\rho}\left(\Delta\right)\int^{\infty}_{0}\mathrm{e}^{\rmi\omega\tau}\mathrm{e}^{\rmi\Delta\tau}\rmd\tau \rmd\Delta. \label{Hijhom}
\end{eqnarray}
To get analytical insight, we consider symmetric transitions, that is to say, equal electric dipoles $\left(\left|\vec{d}_{13}\right|=\left|\vec{d}_{23}\right|=d\right)$ 
and inhomogeneous broadening distributions $\left(G_{13}\left(\Delta\right)=G_{23}\left(\Delta\right)=G\left(\Delta\right)\right)$. Under these assumptions it follows that $H_{13}(\omega)=H_{23}(\omega)=H(\omega)$ and $\eta_{13}=\eta_{23}=\eta$. Therefore the solutions of equations (\ref{fieldeqijfin}) read:
\begin{eqnarray} 
\widetilde{\Omega}^{\rm in}_{\mu\rho}(z,\omega)=\widetilde{\Omega}_{\mu\rho}^{\rm in}(0,\omega)\exp\left[-\frac{\alpha(\omega) z}{2} \right],
\label{solfieldfin}
\end{eqnarray}

\subsection{Backward and Forward Retrieval}

We study now the propagation of the retrieved light pulse in backward and forward directions caused by the sign change of the detunings, after the absorption stage, at time $t=0$. Note that for the backward propagating case, a phase matching operation is also needed for the retrieval which corresponds to a sign change in the field amplitudes. Therefore, by changing $\Delta\rightarrow-\Delta$ and $\Omega^{\rm b,f}_{\mu\rho}\rightarrow-\Omega^{\rm b,f}_{\mu\rho}$ for the backward and $\Delta\rightarrow-\Delta$ for the forward retrieval in equations (\ref{sigma13eq}), equations (\ref{sigma13eq})-(\ref{fieldeqs}) can be solved following the same method as in the absorption stage, using the boundary conditions $\widetilde{\Omega}_{\mu\rho}^{\brnew}(0,\omega)=0$ and $\widetilde{\Omega}_{\mu\rho}^{\brnew}(L,\omega)=0$ for the forward and backward re-emission processes, respectively. By defining
\begin{subequations}\label{FijJij}
\begin{eqnarray}
F_{\mu\rho}(\omega) &=&\int_{-\infty}^{+\infty}G_{\mu\rho}\left(-\Delta\right)
\int_{0}^{\infty}\mathrm{e}^{\rmi{\omega}\tau}\mathrm{e}^{-\rmi{\Delta}\tau}\rmd\tau \rmd\Delta, \label{Fijhom} \notag \\ \\
J_{\mu\rho}(\omega) &=&\int_{-\infty}^{+\infty}G_{\mu\rho}\left(-\Delta\right)
\int_{-\infty}^{+\infty}\mathrm{e}^{\rmi{\omega}\tau}\mathrm{e}^{-\rmi{\Delta}\tau}\rmd\tau \rmd\Delta. \label{Jijhom}  \notag \\ 
\end{eqnarray}
\end{subequations}
the equations for field components associated with each transition are: 
\begin{eqnarray} \label{fieldeq13bout}
	\frac{\partial}{\partial z} \widetilde{\Omega}_{\mu\rho}^{\brnew,\fr}(z,\omega)&=&\pm
	 \left[F_{\mu\rho}(\omega)\widetilde{\Omega}_{\mu\rho}^{\brnew,\fr}(z,\omega)\right. \notag \\
&+& \left.J_{\mu\rho}(\omega)\right. \left.\widetilde{\Omega}_{\mu\rho}^{\rm in}(z,-\omega)\right]\eta_{\mu\rho},
\end{eqnarray}
with $+$ ($-$) for the backward (forward) configurations. The analytical solutions for the output fields read:
\begin{eqnarray}
 \widetilde{\Omega}^{\brnew,\fr}_{\mu\rho}(z,\omega)&=&-\gamma^{\brnew,\fr}(\omega) \widetilde{\Omega}^{\rm in}_{\mu\rho}(0,-\omega),
 \label{solfieldb}
\end{eqnarray}
where we have assumed symmetric transitions and we have defined the backward and forward retrieval coefficients as:
\begin{subequations}
\begin{eqnarray}
\gamma^\brnew(\omega)&=&\frac{J(\omega)}{F(\omega)+H(-\omega)}
 \left(1-\mathrm{e}^{-L\eta \left(F\left(\omega\right)+H\left(-\omega\right)\right)}\right) \label{backCoeff}\notag \\ \\
\gamma^\fr(\omega)&=& z\eta J(\omega) \mathrm{e}^{-z\eta \frac{F(\omega)+H(-\omega)}{2}}\times \notag \\
 &\times& {\rm sinhc}\left(z\eta \frac{F(\omega)-H(-\omega)}{2}\right) \label{forwCoeff}
\end{eqnarray}
\end{subequations}
with ${\rm sinhc}(x)=\sinh (x) /x$ denoting the hyperbolic sinus cardinal function.
When the spectral bandwidth of the pulse is smaller than that of the inhomogeneously broadened transitions,
the functions defined in (\ref{Hijhom}) and (\ref{FijJij}) can be approximated as $F(\omega)\simeq H(-\omega)\simeq J(\omega)/2\simeq\pi/2$. Therefore the absorption coefficient becomes
$\alpha(\omega)=\eta J(\omega)\simeq\eta \pi$ and $\gamma^{\brnew,\fr}(\omega)$ simplify to:
\begin{subequations}
\begin{eqnarray}
\gamma^\brnew&\simeq&(1-\mathrm{e}^{-\alpha L}) {\ \rm and} \label{simpBackCoeff} \\
\gamma^\fr&\simeq&\alpha L \mathrm{e}^{-\alpha L/2}, \label{simpForwCoeff}
\end{eqnarray}
\end{subequations}
with $\alpha L$ being the optical depth.
Equation~(\ref{solfieldb}) shows that for the backward scheme both field components are recovered up to the factor $\gamma^\brnew$, which approaches unity for a large enough medium length. 
This result confirms that ideally both polarization components can be stored and recovered perfectly in the backward direction, thus permitting to build up a quantum memory for a polarization qubit. On the other hand, in the forward retrieval case the quantum memory yields a forward retrieval with a maximum achievable efficiency around $54\%$ as expected from \cite{SangouardPRA07}.

\section{\label{sec:PhaseNoise}Effects of Phase noise}

We consider now the effects of a specific variation in the phase of the atomic coherences. This is propaedeutic to the analysis of the role of the phase noise in the quantum memory, which will be the main goal of this section. Let us consider that, after the complete absorption of the field and before the reversing of the detuning, the phase difference between the atomic levels can be artificially engineered, for instance, by means of 
external electric or magnetic fields. The phase variation of the atomic coherences can be written, in the most general case, as:
\begin{eqnarray}
 \sigma_{ij}\mapsto \mathrm{e}^{\rmi(\phi_i-\phi_j)}\sigma_{ij} \label{atom_ph}
\end{eqnarray}
with $i,j=1,2,3$. This phase variation affects the quantum memory protocol yielding an additional phase to the initial conditions used in the retrieval stage of the protocol. As a consequence, \refeqs{fieldeq13bout} transform as follows:
\begin{eqnarray} \label{fieldeq13fboutphase}
	\frac{\partial}{\partial z} \widetilde{\Omega}_{\mu\rho}^{\brnew,\fr}(z,\omega)&=&
	 \pm \left[F_{\mu\rho}(\omega)\widetilde{\Omega}_{\mu\rho}^{\brnew,\fr}(z,\omega)\right. \notag \\
&+& \left.J_{\mu\rho}(\omega)\mathrm{e}^{\rmi(\phi_\mu-\phi_\rho)}\widetilde{\Omega}_{\mu\rho}^{\rm in}(z,-\omega)\right]\eta_{\mu\rho},
\end{eqnarray}
with $+$ ($-$) corresponding to the backward (forward) retrieval protocol. The solutions of equations(\ref{fieldeq13fboutphase}) read:
\begin{eqnarray}
 \widetilde{\Omega}^{\brnew,\fr}_{\mu\rho}(z,\omega)&=&-\mathrm{e}^{\rmi(\phi_\mu-\phi_\rho)}\gamma^{\brnew,\fr}(\omega) \widetilde{\Omega}^{\rm in}_{\mu\rho}(0,-\omega).
 \label{solb_phi}
\end{eqnarray}
The above expression is identical to \refeqs{solfieldb}, apart from a phase factor. As a consequence, the retrieval efficiency for each field is identical to the standard CRIB protocol. However the additional phase change introduced here affects the fidelity, defined in \refeq{def_fid}. Let us focus on the most interesting case corresponding to the backward retrieval, for which each component of the pulse can be recovered with unit efficiency for large optical depth media (\ie, $\gamma^\brnew(\omega)\rightarrow 1$). Consider a generic polarization qubit at the input:
\begin{eqnarray}
|\psi\rangle=c_{L}|L\rangle + c_{R}|R\rangle,
\label{input}
\end{eqnarray}
where $|L\rangle$ and $|R\rangle$ denote a left- and right- circularly polarization states of a single photon, respectively. In the case of unit efficiency, the output backward field (\ref{solb_phi}) would be:
\begin{eqnarray}
|\psi\rangle=c_{L} \mathrm{e}^{\rmi\phi_L}|L\rangle + c_{R} \mathrm{e}^{\rmi\phi_R}|R\rangle
\label{output}
\end{eqnarray}
being $\phi_L=\phi_3-\phi_1$ and $\phi_R=\phi_3-\phi_2$. From \refeq{def_fid}, the corresponding input-output fidelity reads
\begin{eqnarray}
F=\left| |c_{L}|^2+\mathrm{e}^{\rmi\phi}|c_{R}|^2\right|^2,
\label{fid_phi}
\end{eqnarray}
with $\phi=\phi_1-\phi_2$. Clearly, even if each component of the initial qubit is completely recovered, the additional phase accumulated in levels $1$ and $2$ degrades the fidelity. The latter is instead unaffected by a phase change in the lower level since, in this case, no phase difference is accumulated between the coherences $\sigma_{13}$ and $\sigma_{23}$.

The analysis performed up to here has been referred to a system evolving in the absence of noise. Let us consider now the presence of noise in the phase of the atomic transitions. Two mechanisms can be foreseen that yield phase noise in the atomic transitions during the storage time. First, consider that after the pulse is absorbed, each atom ---apart from evolving according to its free Hamiltonian--- is affected by spurious magnetic and electric fields. This would lead to a non-ideal rephasing after the reversing of the detuning. Second, as it has been discussed in the Introduction, in order to increase the storage time it is customary to temporally transfer the population from the excited unstable level to an auxiliary level endowed with larger coherence time. This procedure is performed via the coupling with light pulses of area $\pi$ \cite{MK,crib,Kraus'06}. Ideally, these pulses should couple to the system without introducing any uncontrolled phase shift to the original atomic transition. However, in any realistic situation, the intensity and phase fluctuations of the light pulses will result in a phase shift, probably different for each atom, when the population is transferred back to the original excited level.
\begin{figure}[t]
{\includegraphics[width=0.8\columnwidth]{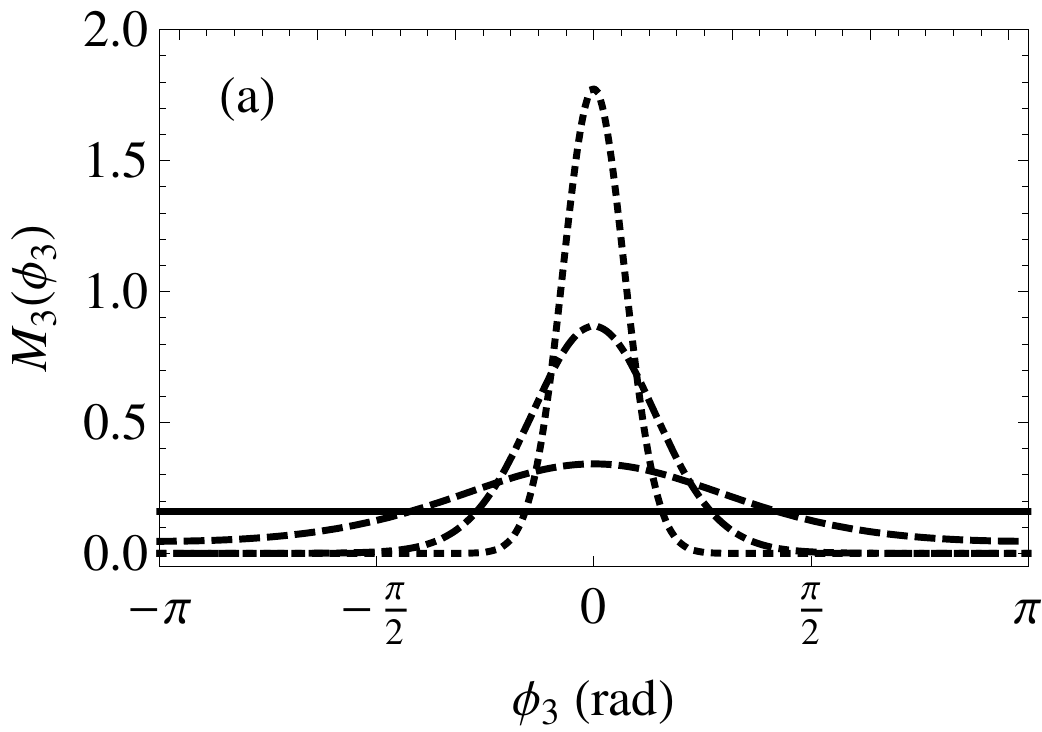}}
{\includegraphics[width=0.8\columnwidth]{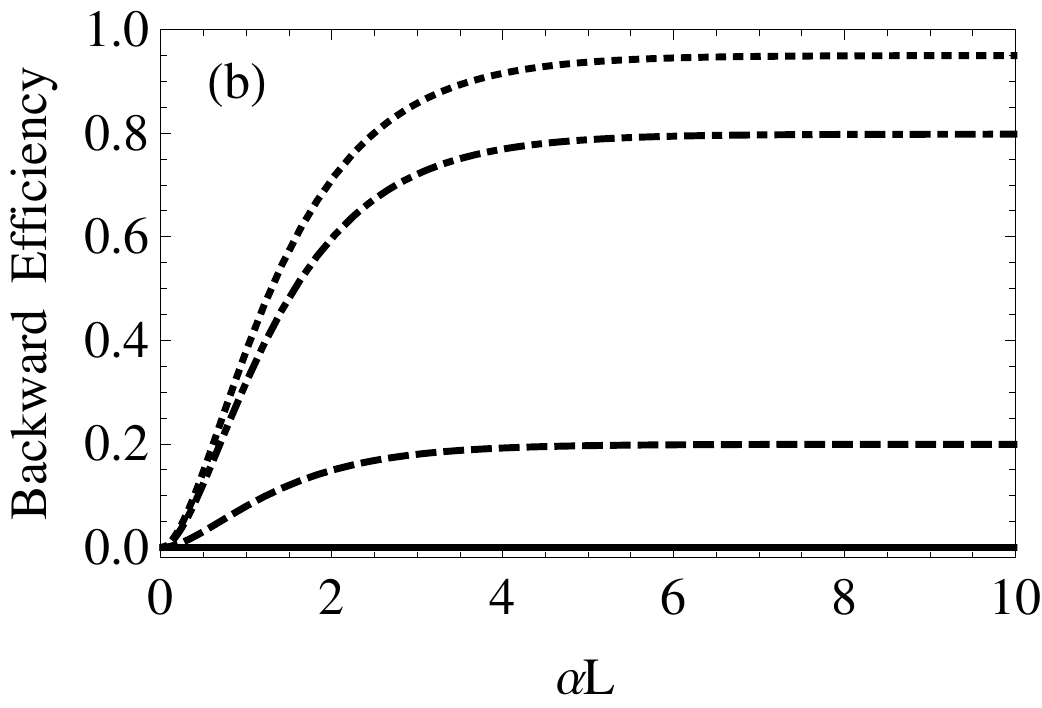}}
{\includegraphics[width=0.8\columnwidth]{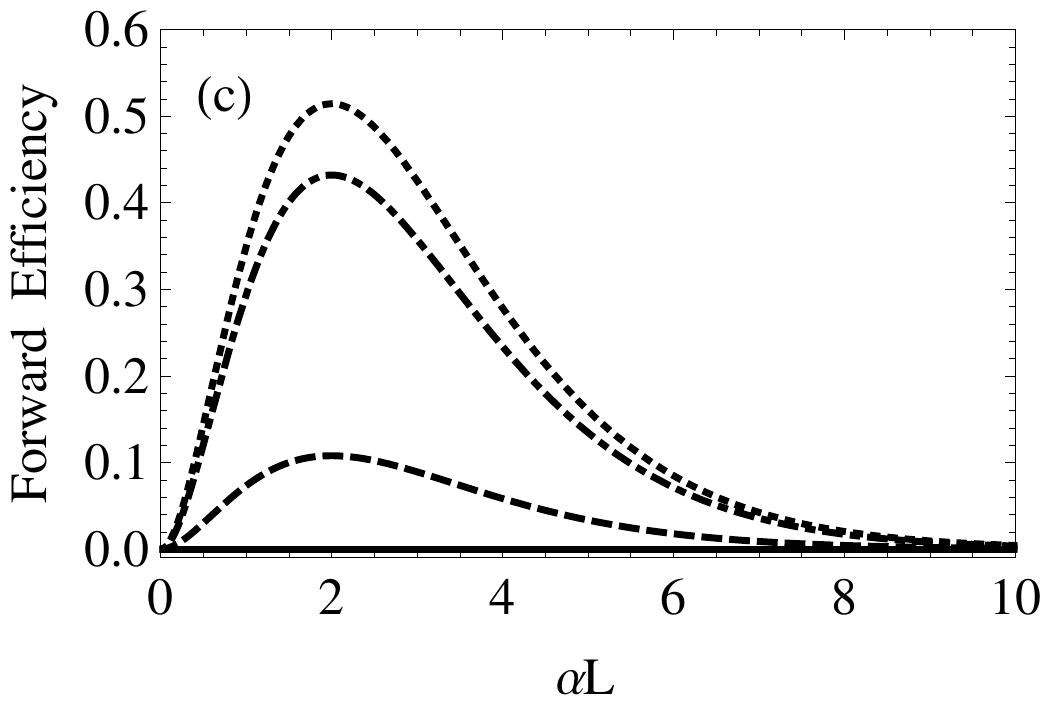}}
\caption{(a) von Mises distribution of the phase of the lower level $\left|3\right\rangle$, for different values of the inverse width distribution: $k_3=0$ (solid line), $k_3=1$ (dashed line), $k_3=5$ (dot-dashed line), $k_3=20$ (dotted line). Memory efficiency for the (b) backward and (c) forward retrieval schemes as a function of the optical depth for different amounts of phase noise affecting the ground level after the absorption of the pulse. The different lines correspond to the same inverse widths of the distribution as in (a).}
\label{f:back_eff_uphnoise}\end{figure}
\begin{figure}[t]
{\includegraphics[width=0.8\columnwidth]{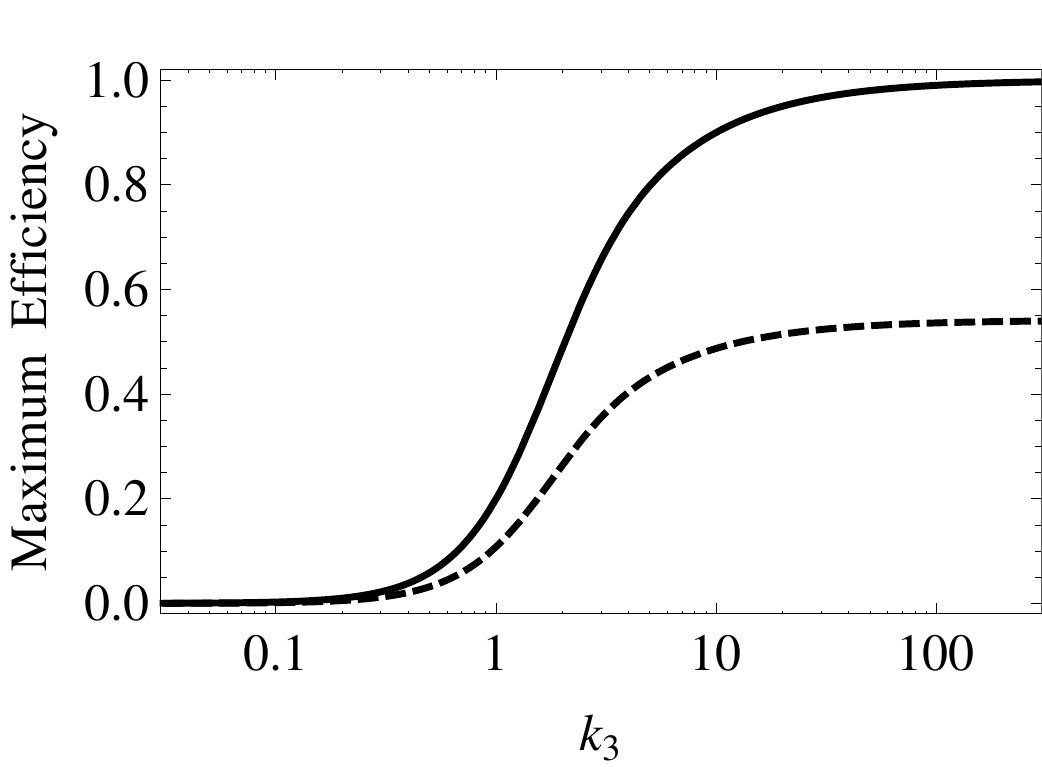}}
\caption{Maximum memory efficiency for the backward (solid line) and the forward (dashed line) retrieval schemes as a function of the inverse width of the phase noise distribution $k_3$.}
\label{f:back_asymeff_uphnoise}
\end{figure}
Both  aforementioned noise mechanisms can be modeled by adding a phase noise to each atomic transition. In practice each atomic coherence is subjected to phase transformations as those given by \refeq{atom_ph}. Since the additional phase is random and centered in zero, we consider a von Mises phase distribution (circular normal distribution), that is to say, a continuous probability distribution on a circle:
\begin{eqnarray}
M_j(\phi_{j})=\frac{\mathrm{e}^{k_{j} \cos\phi_{j}}}{2 \pi I_0(k_{j})}.
\label{vonmises}
\end{eqnarray}
In the definition above $j$ refers to the level affected by the noise, $\phi_{j}$ is the phase angle, $I_0(k_{j})$ denotes the modified Bessel function of zero order, and $k_{j}$ is the inverse width of the phase noise distribution, that is assumed to be centered at zero phase value. For $k_{j}=0$ the distribution is uniform along the interval $[-\pi,\pi]$ and the larger the $k_{j}$, the narrower the noise distribution, and vice versa (see figure \ref{f:back_eff_uphnoise}(a)).
Notice that the actual form of the chosen distribution is however not relevant for our purposes.

In order to introduce the effects of phase noise in equations (\ref{sigma13eq})-(\ref{fieldeqf}), one can proceed as in equation (\ref{atom_ph}), but now an additional averaging over all the possible phases must be performed in the field equations:
\begin{eqnarray} \label{fieldeqsphi3}
\frac{\partial}{\partial z}\Omega^{\brnew,\fr}_{\mu\rho}(z,t) &=&\mp \rmi\eta_{\mu\rho}\int^{\pi}_{-\pi}{\rmd\phi_{1}\rmd\phi_{2}\rmd\phi_{3}}\int^{\infty}_{-\infty}{\rmd\Delta}\left[M_{T}\right.\times \notag \\ 
&& \times\left.G_{\mu\rho}\left(\Delta\right) \sigma_{\mu\nu}^{\brnew,\fr}(z,t,\Delta,\phi_1,\phi_2,\phi_3)\right]. \notag \\
\end{eqnarray}
where $M_T\equiv M_{1}\left(\phi_1\right)M_{2}\left(\phi_2\right)M_{3}\left(\phi_3\right)$. For a generic noise acting upon all the atomic levels, the expressions for the retrieved fields, both in the forward and backward configurations, read as in \refeqs{solb_phi} except for a multiplying factor:
\begin{eqnarray}\label{solf_noise}
 \widetilde{\Omega}^{\brnew,\fr}_{\mu\rho}(z,\omega)&=&-K_{\rho}K_{\mu}\gamma^{\brnew,\fr}(\omega) \widetilde{\Omega}^{\rm in}_{\mu\rho}(0,-\omega), 
\end{eqnarray}
where $K_{\mu}=\int_{-\pi}^{\pi}{\mathrm{e}^{\pm \rmi\phi_{\mu}} M_\mu(\phi_{\mu})\rmd\phi_{\mu}}=I_1(k_\mu)/I_0(k_\mu)$ that increases, monotonously with $k_{\mu}$, from zero to one.

Consider, for instance, that the phase noise acts upon level $\left|3\right\rangle$ only (the case of noise acting on the other levels being similar). From expressions (\ref{solf_noise}), the retrieval efficiencies for the backward and forward modes can be obtained:
\begin{eqnarray} 
 {\rm Eff}^{\brnew,\fr}_{\mu\rho}&=&\left(K_{3}\gamma^{\brnew,\fr}\right)^2, \label{beff_noise}
\end{eqnarray}
where $\gamma^{\brnew}$ and $\gamma^{\fr}$ are defined in (\ref{simpBackCoeff}) and (\ref{simpForwCoeff}), respectively. Notice that the efficiency does not depend on the amplitude components of the initial field, so it is independent of the particular superposition state we are considering.
Focusing in the backward scheme,
we plot in figure \ref{f:back_eff_uphnoise}(b) the retrieval efficiency as a function of the optical depth for different values of the inverse width of the distribution $k_3$.
We see that for large $k_3$ (narrow phase noise distribution)
the efficiency is unaffected by the presence of noise, and a perfect recovery of the input light pulse can be reached for a large enough optical depth. However, as $k_3$ becomes smaller (a wider noise distribution), the efficiency suffers from relevant detrimental effects. In this case, the input pulse cannot be recovered completely, even for a large optical depth. For the forward retrieval scheme, similar results are obtained (see figure \ref{f:back_eff_uphnoise}(c)) with the obvious difference that in this case the recovery efficiency is limited to $\sim54\%$. These results do not contradict the above mentioned fact that both the efficiency and fidelity are unaffected 
by a single specific phase change $\phi_3$ of the ground state (see \refeqs{solb_phi}, and (\ref{fid_phi})).
In fact, when the phase noise is present, the situation is different: since each atom accumulates a different phase during the storage time, the rephasing process, which is a collective phenomenon, can not be achieved properly. 

In order to have a compact assessment of the effect of the phase noise, we plot in figure \ref{f:back_asymeff_uphnoise} the maximum value of the backward (solid line) and forward (dashed line) efficiency as a function of the noise parameter $k_{3}$, for the case where the phase noise is only included in level $\left|3\right\rangle$. 
We can see that the memory efficiency in both backward and forward schemes degrades substantially as the phase noise distribution becomes wider. Note that the actual values of the noise depend on the  chosen model, however the qualitative behavior is a general feature.

In order to obtain a complete picture of the retrieval process in the presence of noise, the input-output fidelity has to be considered too. As indicated, whenever noise is present the retrieval efficiency is reduced with respect to the ideal case, being below the unity also in the backward scheme. This implies that the output state $\varrho^{\brnew,\fr}$ is not pure any more. The expression of $\varrho^{\brnew,\fr}$ can be obtained, however, simply by modeling the non-ideal memory with an ideal one preceded by a beam-splitter with transmittivity equal to the efficiency of the memory. The reflected beam is then traced out. In this way, one obtains that the fidelity is equal to the retrieval efficiency, in the case of polarization qubits.

\FloatBarrier

\section{\label{sec:Numerics}Numerical Results}

Analytical results reported in previous sections are based on the weak field approximation in which the temporal dynamics of the atomic populations and the two-photon coherence $\sigma_{12}$ was neglected. In this section we verify the validity of the analytical approach by performing numerical integration of the full optical-Bloch equations (\ref{sigma13eq}-\ref{fieldeqs}) with a finite difference method beyond this approximation. Atoms with closely spaced frequencies are grouped 
in equidistantly spaced frequency classes 
(labeled by an integer number $n$) with 
central frequencies $\omega_{13}^{\left( n \right)}=\omega_{23}^{\left( n \right)}$.
At each spatial point $z$ the inhomogeneous spectral distribution (provided by the transverse broadening)
of the frequency classes is given by the function $G\left(\Delta^{\left( n \right)}\right)$ which, for simplicity, is assumed to be a rectangular distribution of width much larger than the spectral width of the input pulse.

The calculation scheme is performed in the following steps: (i) From the temporal profile of the incident field pulse $\Omega_{\mu\rho}(z,t)$, we calculate the temporal evolution of the density matrix at each spatial point $z$ and for each frequency class by integration of equations (\ref{sigma13eq}) with time step $\Delta t$. (ii) This temporal evolution given by $\sigma_{\mu\rho}\left(z,t,\Delta^{\left( n \right)}\right)$ is used in equations (\ref{fieldeqs}) to propagate the field pulses between the spatial points $z$ and $z + \Delta z$. Further, steps (i) and (ii) are iterated. Temporal $\Delta t$ and spatial $\Delta z$ steps are chosen small enough and the number of	frequency classes large enough to assure the convergence of the calculations.

\begin{figure} 
{\includegraphics[height=0.8\columnwidth]{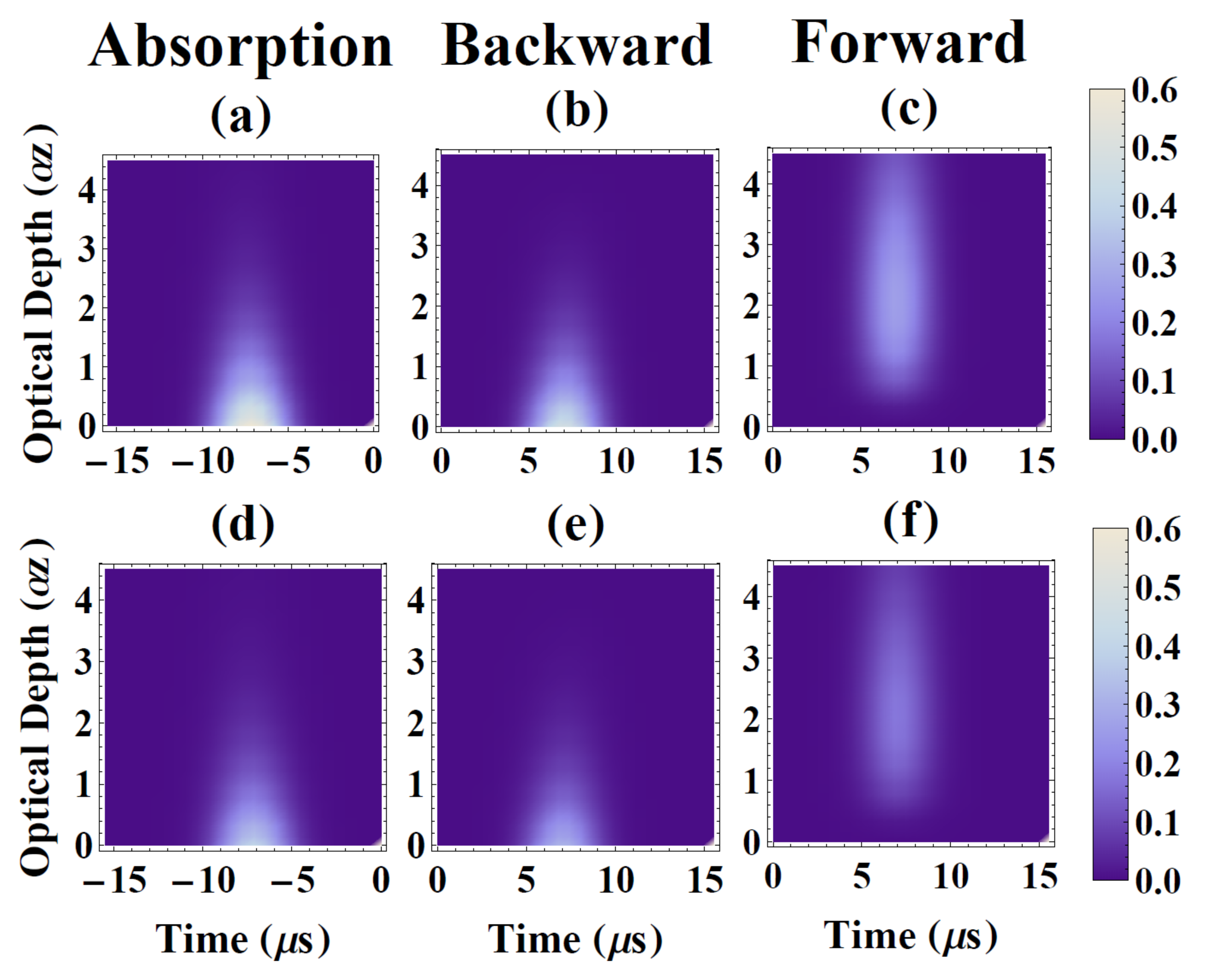}}
\caption{Contour plots of the spatio-temporal dynamics of the normalized intensities of the two circular polarization components of the field 
$I_{13}(z,t)$ (a-c) and $I_{23}(z,t)$ (d-f) in an inhomogeneously broadened $V$-system with 
phase noise distribution of width $k_{3}=5$ in the lower level $\left|3\right\rangle$.
Figures (a,d) show the absorption stage, 
while (b,e), and (c,f) correspond to the backward and forward retrieval schemes, respectively.}
\label{Numerics_V_G3}
\end{figure}

During the absorption stage, the pulse propagation is initiated at $z=0$, where the density matrices of all atom frequency classes are known. As in the analytical approach, we assume that initially all atoms are in their ground level, {\it i.e.}, $\sigma_{33} \left(z,t=-15\mu s \right) = 1$. The input field pulses at $z=0$ are assumed to have Gaussian temporal profiles and we define the normalized intensities of each component as
\begin{eqnarray} 
	I_{\mu\rho}(z,t)= \frac{\left| \Omega_{\mu\rho}(z,t) \right|^{2}}{\left|\Omega_{13}(0,t_{c}) \right|^{2}+\left|\Omega_{23}(0,t_{c}) \right|^{2}}\;, \label{norm_int}
\end{eqnarray}
where $\mu=1,2$, $\rho=3$ and $t_{c}$ is the time corresponding to the peak of the pulse. The spatio-temporal dynamics of $I_{13}(z,t)$ and $I_{23}(z,t)$ during the absorption stage of the protocol is shown in figures \ref{Numerics_V_G3}(a) and (d), respectively. The initial intensities of the components are taken as $I_{13}(0,t_{c})=0.6$ and $I_{23}(0,t_{c})=0.4$. Note that the absence of field at the output surface, corresponding to an optical depth $\alpha z=\alpha L=4.5$ with $L$ being the length of the medium, evidences the efficient absorption of both field components.

For the simulation of the forward and backward retrieval protocols we reverse the sign of the detuning for each frequency class $\Delta^{\left(n\right)} \rightarrow -\Delta^{\left(n\right)}$.
For the backward protocol, we additionally change the sign of the field to implement the phase matching operation. After this step, the phase noise is introduced by imposing at each spatial position a phase distribution in the atomic coherences of each frequency class, $n$, given by equation (\ref{vonmises}), {\it i.e.}, $\sigma_{\mu \nu}^{\left( n \right)} \left( z,t=0 \right) =
\left| \sigma_{\mu \nu}^{\left( n \right)} \left( z,t=0 \right) \right| \mathrm{e}^{\rmi \phi_{\mu \nu 0}^{\left( n \right)}} $. The ensemble average $\widetilde{\sigma}_{\mu \nu}^{\left( n \right)} = \left\langle \sigma_{\mu \nu}^{\left( n \right)} \right\rangle$ is used as the initial density matrix element for each frequency class at the retrieval stage. During the retrieval stage the field propagation is evaluated by means of equations (\ref{sigma13eq}) and (\ref{fieldeqs}) from the surface $\alpha z=0$ ($\alpha z=\alpha L$) to the surface $\alpha z=\alpha L$ ($\alpha z=0$) for the forward (backward) protocol. The spatio-temporal dynamics of the retrieved field intensities, $I_{13}$ and $I_{23}$, in the presence of a phase noise distribution in the lower level $\left|3\right\rangle$ of inverse width $k_{3}=5$ are shown in figures \ref{Numerics_V_G3}(b) and (e) for the backward retrieval and in figures \ref{Numerics_V_G3}(c) and (f) for the forward retrieval protocols.
 
For the parameters used in our simulations we have numerically observed that the two-photon coherence ($\sigma_{12}$),
the population of the excited levels ($\sigma_{11}$, $\sigma_{22}$), and the phase difference between the field components did not exceed $10^{-6}$ during the whole dynamics, validating the weak field approximation used to derive the analytical solutions in Section~\ref{sec:ideal case}. By comparing the absorption of the field components (figures \ref{Numerics_V_G3}(a) and (d)) with the corresponding retrieved fields (figures \ref{Numerics_V_G3}(b),(c) and (e),(f), respectively), we note that, although the efficiency of the re-emission is lower than in the ideal case where no phase noise is present, the decreasing on the retrieval intensities is the same for both components.

We have checked that beyond this particular case, the propagation dynamics and the maximal retrieval efficiencies acquired from the numerical simulations for both forward and backward protocols are in good agreement with the analytical results, confirming the validity of the approximations that we assumed in the analytical approach. In particular, for the parameter values of figure \ref{Numerics_V_G3}, the maximum backward and forward numerically evaluated efficiencies read ${\rm Eff}^{\brnew}_{\rm (num)}=0.79$ and ${\rm Eff}^{\fr}_{\rm (num)}=0.43$, respectively that match the analytical results shown in figure \ref{f:back_eff_uphnoise} for $k_3=5$ (dot-dashed line).

\section{\label{sec:Conclusions}Conclusions}

In this paper we have shown that optical quantum memories based on $V$-type three-level atoms can be used to store and retrieve polarization qubits by means of the 
CRIB technique without the need of two separate quantum memories or spatially splitting the input state, hence presenting experimental compactness. In the weak field regime,
the optical-Bloch equations for the three-level atoms decouple into two sets of equations describing the evolution of 
the two circular polarization components of the field resulting in two independent two-level quantum memories. In this scenario, we have analytically studied the effect of an extra phase in the atomic levels. We have shown that when the phase variations affect equally all the atoms of the sample, the fidelity is degraded whereas the efficiency for each polarization component of the pulse is unaffected. In contrast, when we consider an inhomogeneously distributed phase noise for the different levels,
the rephasing process during the retrieval stage is not properly achieved and both the efficiency and the fidelity are degraded.
In addition, we have numerically checked the assumptions performed in the analytical approach, by numerically integrating the full optical-Bloch equations.
The good agreement between the numerical and analytical results demonstrates the validity of the analytical results obtained under the weak field approximation.

Along this work we have proposed the implementation of a quantum memory for qubits in a superposition state of left and right circular polarizations in $V$-type three-level atoms. In the original CRIB proposal \cite{Kraus'06} rare-earth-ion-doped (REID) crystals were suggested as possible candidates to implement quantum memories due to its well-known energy level structure and long coherence times at low temperatures \cite{spinCoh}. However, although $V$-type three-level configurations have been implemented in REID crystals (see for instance \cite{Beil'08} for a EIT experiment in $\rm Pr^{3+}:Y_{2}SiO_{5}$), the optical transitions in this kind of systems are not polarization selective, so the only way to address them separately is to use a source whose bandwidth is smaller than the ground-state sublevel splitting. This restriction is an important drawback for the implementation of our proposal in REID, where a system with specific selection rules is needed to absorb the left and right circular polarization components of the photonic qubit. Nevertheless, atomic vapours do not suffer from these limitations and could be good candidates to implement a polarization qubit three-level memory. In particular, regarding photon echo quantum memories, warm \cite{Hosseini'09} and cold \cite{Sparkes'10} atoms of $\rm ^{87}Rb$ have been shown to be useful for $\Lambda$-type GEM implementations, with reasonable high efficiencies and storage
times.

Finally, we are presently investigating the implementation of optical quantum memories based on CRIB for $\Lambda$-type three-level atoms. In this last case, the atomic coherence between the two ground states plays an important role in both the absorption and retrieval stages and, in general, the system does not resemble two independent two-level quantum memories. The dynamics of this system, being  much more involved and opening new scenarios for the manipulation of polarization qubits, is out of the scope of the present paper and will be discussed elsewhere.

\begin{acknowledgments}

A.F. gratefully acknowledges C. Ottaviani for insightful comments and suggestions.We acknowledge support from the Spanish Ministry of
Science and Innovation under Contract No. FIS2008-02425,
No. HI2008-0238, and No. CSD2006-00019 (Consolider
project Quantum Optical Information Technologies), from the
Catalan government under Contract No. SGR2009-00347.
\end{acknowledgments}


\bigskip

\begin{thebibliography}{99}

\bibitem{nature_review} For a recent review see Lvovsky A I, Sanders B C and Tittel W 2009 {\it Nature Photonics} {\bf 3} 706

\bibitem{Zhao'09} Zhao R, Dudin Y O, Jenkins S D, Campbell C J, Matsukevich D N, Kennedy T A B and Kuzmich A 2009 {\it Nature Physics} {\bf 5} 100

\bibitem{Riedmatten'10} de Riedmatten H , Nature Photonics {\bf 4}, 206 (2010).

\bibitem{Reim'10}
Reim K F, Nunn J, Lorenz V O, Sussman B J, Lee K C, Langford N K, Jaksch D and Walmsley I A 2010 {\it Nature Photonics} {\bf 4} 218

\bibitem{Saglamyurek'11} Saglamyurek E, Sinclair N, Jin J, Slater J A, Oblak D, Bussie\`eres F, George M, Ricken R, Sohler W and Tittel W 2011 {\it Nature} {\bf 469} 512

\bibitem{Clausen'11} Clausen C, Usmani I, Bussi\`eres F, Sangouard N, Afzelius M, de Riedmatten H and Gisin N 2011 {\it Nature} {\bf 469} 508

\bibitem{LongdellNews'11} Longdell J 2011 {\it Nature} {\bf 469} 475

\bibitem{klm} Knill E, Laflamme R and Milburn G J 2001 {\it Nature} {\bf 409} 46;
Kok P, Munro W J, Nemoto K, Ralph T C, Dowling J P and Milburn G J 2007 {\it Rev. Mod. Phys.} {\bf 79} 135

\bibitem{kimble} Kimble H J 2008 {\it Nature} {\bf 453} 1023

\bibitem{qrep} Briegel H J, D\"{u}r W, Cirac J I and Zoller P 1998 {\it Phys. Rev. Lett.} {\bf 81} 5932; 
Chaneli\`ere T, Matsukevich D N, Jenkins S D, Lan S-Y, Kennedy T A B and Kuzmich A 2005 {\it Nature} {\bf 438} 833;
Eisaman M D, Andr\'e A, Massou F, Fleischhauer M, Zibrov A S and Lukin M D 2005 {\it Nature} {\bf 438} 837; 
Sangouard N, Simon C, Min\'a\ifmmode \check{r}\else \v{r}\fi{} J, Zbinden H, de Riedmatten H and Gisin N 2007 {\it Phys. Rev. A}  {\bf 76} 050301(R);
Sangouard N, Simon C, Zhao B, Chen Y-A, de Riedmatten H, Pan J-W and Gisin N 2008 {\it Phys. Rev. A} {\bf 77} 062301;
Sangouard N, Simon C, de Riedmatten H and Gisin N 2011 {\it Rev. Mod. Phys} {\bf 83} 33

\bibitem{polmem_ensembles} Matsukevich D N and Kuzmich A 2004 {\it Science} {\bf 306} 663; 
Choi K, Deng H, Laurat J and Kimble H 2008 {\it Nature (London)} {\bf 452} 67; 
Chen Y-A, Chen S, Yuan Z-S, Zhao B, Chuu C-S, Schmiedmayer J and Pan J-W 2008 {\it Nature Phys.} {\bf 4} 103; 
Tanji H, Ghosh S, Simon J, Bloom B and Vuleti\'{c} V 2009 {\it Phys. Rev. Lett} {\bf 103} 043601; 
Cho Y-W and Kim Y-H 2010 {\it Opt. Express} {\bf 18} 25786;
Choi K S, Goban A, Papp S B, van Enk S J and Kimble H J 2010 {\it Nature} {\bf 468} 412

\bibitem{polmem_solid} Staudt M U, Afzelius M, de Riedmatten H, Hastings-Simon S R, Simon C, Ricken R, Suche H, Sohler W and Gisin N 2007 {\it Phys. Rev. Lett.} {\bf 99} 173602

\bibitem{Specht'11} Specht H P, N\"olleke C, Reiserer A, Uphoff M, Figueroa E, Ritter S and Rempe G {\it arXiv:quant-ph/1103.1528v1}

\bibitem{Carreño'10} Carre\~{n}o F and Ant\'on M A 2010 Opt. Commun. {\bf 283} 4787

\bibitem{peqm_review} Tittel W, Afzelius M, Chaneli\`ere T, Cone R L, Kr\"{o}ll S, Moiseev S A and Sellars M 2010 {\it Laser \& Photon. Rev.} {\bf 4} 244;
Moiseev S A and Tittel W {\it arXiv:quant-ph/0812.1730v1}

\bibitem{MK} Moiseev S A and Kr\"{o}ll S 2001 {\it Phys. Rev. Lett.} {\bf 87} 173601

\bibitem{crib} Nilsson M and Kr\"{o}ll S 2005 {\it Opt. Commun.} {\bf 247} 393

\bibitem{Kraus'06} Kraus B, Tittel W, Gisin N, Nilsson M, Kr\"oll S and Cirac J I 2006 {\it Phys. Rev. A} {\bf 73} 020302(R)

\bibitem{SangouardPRA07} Sangouard N, Simon C, Afzelius M and Gisin N 2007 {\it Phys. Rev. A} {\bf 75} 032327

\bibitem{Longdell'08} Longdell J J, H\'etet G, Lam P K and Sellars M J 2008 {\it Phys. Rev. A} {\bf 78} 032337

\bibitem{GEM} Alexander A L, Longdell J J, Sellars M J and Manson N B 2006 {\it Phys. Rev. Lett.} {\bf 96} 043602; 
Moiseev S A and Arslanov N M 2008 {\it Phys. Rev. A} {\bf 78} 023803;
H\'etet G, Longdell J J, Alexander A L, Lam P K and Sellars M J 2008 {\it Phys. Rev. Lett.} {\bf 100} 023601;
H\'etet G, Longdell J J, Sellars M J, Lam P K and Buchler B C 2008 {\it Phys. Rev. Lett.} {\bf 101} 203601;
Lauritzen B, Min\'a\ifmmode \check{r}\else \v{r}\fi{} J, de Riedmatten H, Afzelius M, Sangouard N, Simon C and Gisin N 2010 {\it Phys. Rev. Lett.} {\bf 104} 080502.
 
\bibitem{Sparkes'10} Sparkes B M, Hosseini M, H\'etet G, Lam P K and Buchler B C 2010 {\it Phys. Rev. A} {\bf 82} 043847

\bibitem{HetetOptLett'08} H\'etet G, Hosseini M, Sparkes B M, D.~Oblak, Lam P K and Buchler B C 2008 {\it Opt. Lett.} {\bf 33} 2323

\bibitem{Hosseini'09} Hosseini M, Sparkes B M, H\'etet G, Longdell J J, Lam P K and Buchler B C 2009 {\it Nature} {\bf 461} 241

\bibitem{afc} de Riedmatten H, Afzelius M, Staudt M U, Simon C and Gisin N 2008 {\it Nature} {\bf 456} 773; 
Afzelius M, Simon C, de Riedmatten H, and Gisin N 2009 {\it Phys. Rev. A}  {\bf 79} 052329;
Bonarota M, Le Gou\"et J-L and Chaneli\`ere T {\it arXiv:quant-ph/1009.2317v2}; 
Afzelius M, Usmani I, Amari A, Lauritzen B, Walther A, Simon C, Sangouard N, Min\'a\ifmmode \check{r}\else \v{r}\fi{} J, de Riedmatten H, Gisin N and S.~Kr\"oll 2010 {\it Phys. Rev. Lett.} {\bf 104} 040503; 
Usmani I, Afzelius M, de Riedmatten H and Gisin N 2010 {\it Nat. Commun.} {\bf 1:12}; 
Lauritzen B, Min\'a\ifmmode \check{r}\else \v{r}\fi{} J, de Riedmatten H, Afzelius M and Gisin N 2011 {\it Phys. Rev. A} {\bf 83} 012318

\bibitem{spinCoh} Longdell J J, Fraval E, Sellars M J and Manson N B {\it arXiv:quant-ph/0506233v1};
Fraval E, Sellars M J and Longdell J J 2004 {\it Phys. Rev. Lett.} {\bf 92} 077601;
Fraval E, Sellars M J and Longdell J J 2005 {\it Phys. Rev. Lett.} {\bf 95} 030506

\bibitem{Beil'08} Beil F, Klein J, Nikoghosyan G and Halfmann T 2008 {\it J. Phys. B: At. Mol. Opt. Phys.} {\bf 41} 074001.

\end{thebibliography}
\end{document}